\documentclass[review]{elsarticle}

\usepackage{lineno,hyperref}
\modulolinenumbers[5]

\journal{Journal of \LaTeX\ Templates}

\usepackage[T1]{fontenc}
\usepackage{lmodern}
\usepackage{microtype}
\usepackage{color}
\usepackage{tabularx}
\usepackage{booktabs}
\usepackage{multirow}
\usepackage[
	format=hang,
	indention=-1.0cm,
	justification=RaggedRight,
	font=small
]{caption}
\usepackage{graphicx} 
\usepackage{dcolumn} 
\usepackage{bm} 
\usepackage[version=3]{mhchem}
\usepackage{stackrel}

\usepackage{verbatim}   
\usepackage{subfigure}  
\usepackage{hyperref}   
\usepackage{float}
\usepackage{amsfonts} 
\usepackage[utf8]{inputenc}
\usepackage[english]{babel}
\usepackage{calc}

\graphicspath{{Figures/}} 
\newlength\myheight
\newlength\mydepth
\settototalheight\myheight{Xygp}
\newcommand{\be}{\begin{equation}}
\newcommand{\ee}{\end{equation}}
\newcommand{\bea}{\begin{eqnarray}}
\newcommand{\eea}{\end{eqnarray}}

\addto\captionsenglish{}

\newcommand{\D}{\mathrm{d}}
\newcommand{\E}{\mathrm{e}}

\begin{document}

\title{The generality of transient compartmentalization and its associated error thresholds}



\author[mymainaddress,mysecondaryaddress]{Alex Blokhuis \corref{mycorrespondingauthor}}
\cortext[mycorrespondingauthor]{Corresponding author}
\ead{alex{\_}blokhuis@hotmail.com}

\author[mysecondaryaddress]{Philippe Nghe}

\author[mythirdaddress]{Luca Peliti}

\author[mymainaddress]{David Lacoste}

\address[mymainaddress]{Gulliver Laboratory, UMR CNRS 7083, PSL Research University, 
ESPCI, 10 rue Vauquelin, F-75231 Paris, France}
\address[mysecondaryaddress]{Laboratory of Biochemistry, PSL Research University, 
ESPCI, 10 rue Vauquelin, F-75231 Paris, France}
\address[mythirdaddress]{SMRI, 00058 Santa Marinella (RM), Italy}

\date{\today}

\date{\today} 

\begin{abstract}
Can prelife proceed without cell division? A recently proposed mechanism suggests that 
transient compartmentalization could have preceded cell division in prebiotic scenarios.
Here, we study transient compartmentalization dynamics in the presence of mutations and noise in replication, 
as both can be detrimental the survival of compartments. 
Our study comprises situations where compartments contain uncoupled autocatalytic reactions feeding on a common resource, 
and systems based on RNA molecules copied by replicases, following a recent experimental study. 

Using the theory of branching processes, we show analytically that two regimes are possible.
In the diffusion-limited regime, replication is asynchronous which leads to a large variability in the composition of compartments.
In contrast, in a replication-limited regime, the growth is synchronous and thus the compositional variability is low. 
Typically, simple autocatalysts are in the former regime, while polymeric replicators can access the latter.

For deterministic growth dynamics, we introduce mutations that turn functional 
replicators into parasites. We derive the phase boundary separating coexistence or parasite 
dominance as a function of relative growth, inoculation size and mutation rate. We show that transient 
compartmentalization allows coexistence beyond the classical error threshold, above which the parasite dominates. 
Our findings invite to revisit major prebiotic transitions, notably the transitions towards cooperation, complex polymers and cell division.

\end{abstract}


\maketitle

\thispagestyle{empty}

\section{Introduction}

\label{TC1}

Compartments play a central role in many biological processes of cells, 
in particular in organelles such as the ER or in the Golgi apparatus \cite{Jaiman2018}. Cells use 
compartments to organize chemical reactions in space: compartments eliminate
the risk of losing costly catalysts which are essential for biochemical 
reactions, they also accelerate chemical reactions, while reducing  
the risk of crosstalk due to other side reactions. 

In the early 20th century, Oparin suggested that membrane-less compartments, which he called coacervates, could have played a central role in the origin of life \cite{Oparin1952}. 
Recently, this idea has resurfaced, after such compartments had been found in organisms, e.g. P-granules in C. Elegans embryos \cite{Brangwynne2009}. These membrane-less compartments represent  a particularly interesting example of active phase separation \cite{Zwicker2017}, and for this reason many groups are trying to synthetize them \cite{Brinke2018,Nakashima2018}.

After the discovery of the structure of DNA, the  
coacervates scenario for the origin of life got less popular, 
and replication scenarios became the new paradigm \cite{Higgs2015,Dyson1985}. 
In the sixties, Spiegelman showed that RNA could be replicated by an enzyme called $Q\beta$ RNA replicase, 
in the presence of free nucleotides and salt \cite{SPIEGELMAN1965}. 
After a series of serial transfers, he observed the appearance of shorter RNA polymers, 
which he called parasites. Typically, these parasites are non-functional molecules 
which replicate faster than the 
RNA polymers introduced at the beginning of the experiment. 
In 1971, Eigen conceptualized this observation by proving theoretically 
that for a given accuracy of replication and a relative fitness of parasites, 
there is a maximal genome length that can be maintained without errors \cite{Eigen1971}. This 
result led to the following paradox: to be a functional replicator, a molecule must be long enough. However, if it is long, 
it cannot be maintained since it will quickly be overtaken by parasites. This error threshold can be interpreted in terms of the information, which needs to be maintained and passed to the next generation. In this context, there are two distinct aspects of information : i) digital ({\it i.e} discrete discontinuous) information, stored in the sequence of functional information-carrying molecules and ii) compositional (continuous) information contained in the compositions of compartments \cite{Szathmary2006}. 
Metabolism-first scenarios put an emphasis on the latter while genetics-first scenarios put an emphasis on the former. Whether one favors one or the other, this puzzle is considered to be a key question in the Origins of life \cite{Smith1995,Takeuchi2012}.

\begin{figure}
\centering
\includegraphics[scale=1.55]{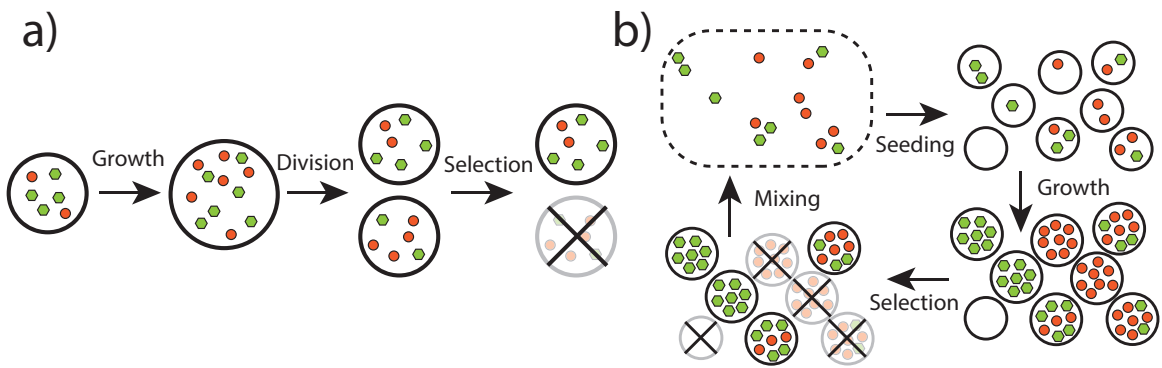} 
\caption{A sketch of a) the stochastic corrector model, b) transient compartmentalization. Both exhibit growth, selection and noisy inoculation of new compartments. } 
\label{fig: tc}
\end{figure}

In the eighties, a theoretical solution to this puzzle was proposed, the Stochastic corrector model~\cite{Szathmary1987} \cite{Grey1995} (see Fig \ref{fig: tc}) inspired by ideas of group selection \cite{Wilson1975}. In the Stochastic corrector model, small groups of replicating molecules grow in a deterministic way in compartments, to a fixed final size 
called the carrying capacity. Then, the compartments are divided and their contents are stochastically partitioned between the two daughter compartments. 
Thanks to the variability introduced by this stochastic division, and to selection acting on compartments, 
a coexistence is possible between replicators and parasites despite the difference in their growth rates. 
The efficiency of this mechanism depends critically on the noise in the transmission of the parent composition (stochastic division) to daughter cells, which is controlled by the carrying capacity.
If the carrying capacity is large, there is not enough variation for group selection to act on. 
If it is too small, frequent random loss of replicators 
will lead to extinction \cite{Grey1995}.

Transient compartmentalization is schematically depicted in Fig \ref{fig: tc}. 
This mechanism was first proposed and experimentally tested by Matsumura et at. \cite{Matsumura2016}. This work inspired us to formulate a general framework 
for transient compartmentalization \cite{Blokhuis2018}.
In its initial formulation, it shared some important features 
with early versions of the Stochastic corrector: \cite{Szathmary1987} 
there is compositional variability in the inoculation step in which molecules from a large pool are used to stochastically seed compartments, then growth is deterministic and selection is performed on a compartment level. 
An essential difference however comes from the mixing step, which does not require cell division. 

Cell division (or more precisely protocell division) does not happen spontaneously.
It requires particular machinery or mechanisms, which may not have been there from the start.  Vesicles are typically impermeable to important biomonomers (nucleotides, amino acids) \cite{Stillwell1976}, which means polymeric replicators in  compartments require some extra features to become sustainable. 
In contrast to this, in transient compartmentalization, resources are encapsulated at the start of each new cycle. Transient compartmentalization can proceed via fluctuations in the environment due for instance to day-night cycles. It is a much more primitive selection mechanism as compared with the Stochastic corrector. 
This simplicity makes the mechanism plausible and general and provides an evolutionary means towards the development of more sophisticated selection mechanisms. As such, transient compartmentalization may have preceded cell-division.

In order to demonstrate this point, we introduced a general class of multilevel selection based on a transient compartmentalization dynamics devoid of compartments division \cite{Blokhuis2018}. It contains a maturation stage, 
in which the contents of a compartment can grow in isolation from compounds present in other compartments. At the end of this growth phase, the contents of compartments are mixed back together. The composition of a compartment may enhance its survival rate: E.g. some compounds may stabilize a compartment (e.g. a vesicle), chelate degradative catalysts (e.g. \ce{Mg^{2+}} ions), buffer a desired chemical environment, improve the influx of metabolites etc. 
The reverse can also be true: compounds may destabilize a compartment, degrade metabolites, catalyze harmful side reactions, harness all replication machinery and so forth. While this may lead to rich and complex phenomena, the effect on survival can often be encoded by a composition-dependent selection function $f(\bar{x})$. By deriving general results for large classes of selection functions, we thus describe a large variety of scenarios.

This class of multilevel selection describes several mechanisms proposed in scenarios for the origins of life, based on various types of compartments (e.g. lipid vesicles \cite{vesicles}, pores \cite{kreysing2015heat,Baaske2007}, inorganic compartments \cite{hydrovent}, coacervates \cite{Oparin1952, Zwicker2017} or aerosols \cite{Dobson2000}) or various protocols of transient compartmentalization \cite{Damer2015,Furubayashi2018}. Particularly exciting is the recent experiment that demonstrated the mechanism for the first time, in which small droplets containing RNA in a microfluidic device \cite{Matsumura2016} were used as compartments. In this experiment, a catalytic RNA was used as a proxy for a functional species/ functional replicator in competition with a nonactive parasite. We have used this system as a model to illustrate the theory of transient compartmentalization\cite{Blokhuis2018}.

The related issue of cooperation between producers and non-producers has been discussed before \cite{Chuang2009}. 
Spatial clustering can lead to similar effects as compartmentalization in 
favoring the survival of cooperating replicators \cite{Tupper2017, Kim2016}.
These ideas were combined in a recent study of a population of individuals 
growing in a large number of compartmentalized habitats, called demes \cite{Geyrhofer2018}. 
Another recent related study on transient compartmentalization quantifies co-encapsulation effects in the context of directed evolution experiments \cite{Zadorin2017}.  

In this paper, we go beyond the analysis carried out in our previous work \cite{Blokhuis2018}
by including the effect of mutations and noise in the growth dynamics of encapsulated populations. By describing mutations, we can derive asymptotic analytical results for the error thresholds in transient compartmentalization. This allows to make the original claim of the stochastic corrector model even stronger: group selection by itself can overcome the error threshold, even without cell division. Another motivation of including mutations comes from experiments, since mutations play a role in the RNA droplet experiment \cite{Matsumura2016} which inspired us. This experiment should be interpreted as a valuable model system to illustrate group selection and chemistry, not as an exact instance of a prebiotic scenario. Our error thresholds are by no means limited to RNA-world scenarios.

The motivation of discussing noise in the growth dynamics in more detail comes from the realization that
replication is inherently stochastic when a small number of replicators are 
present in compartments, which corresponds to the case of the RNA droplet experiment \cite{Matsumura2016}. Naively, one could expect the deterministic approach \cite{Blokhuis2018} to fail in this case, but our more detailed analysis shows that this does not happen for polymer replicators. 
The explanation is that polymer replication involves a succession of a large and well defined number of independent rate-limiting steps, and as a result, the noise in the replication time becomes very small. In Sec \ref{subsec: repmod}, we prove this result, and we also consider the alternative case of a single rate-limiting step for which replication can become very noisy. This agrees with recent works \cite{Houchmandzadeh2018, Michaels2018}, which showed that giant fluctuations are predicted in the final population composition of stochastically growing populations.
 
A similar behavior is expected for uncoupled autocatalytic reactions which can be described with a single replication step, and therefore also present large fluctuations.
Such effects should have a major impact on composition-dependent selection. Indeed, it was shown that a stochastic corrector needs noise to be just large enough for group selection to be efficient, but small enough to mitigate the risk of yielding non-sustainable daughter compartments \cite{Grey1995}. Giant fluctuations in replication would strongly promote extinction in such a mechanism. Molecular replicators that wish to afford more advanced modes of selection are expected to adapt their growth accordingly. 

These results can be interpreted using the distinction between digital (sequence) and compositional information introduced earlier. In small-molecule autocatalytic networks and in 
GARD \cite{Segre1998,Lancet2018}, there are no information-carrying polymers and no template replication, the information is purely analogue. In this case, the control over compositional information is poor, due to the giant fluctuations mentioned above. 
In contrast, when digital information is used thanks to template replication, an improved control over the compositions is gained, because fluctuations in composition are reduced as shown in our model. 
In this case, the loss of the analogue compositional information in transient compartmentalization is compensated by a preservation of  information in a digital form in the sequence of functional molecules thanks to selection \cite{SPIEGELMAN1965}.





In order to emphasize the generality of our approach, we have extended 
in Sec. \ref{sec:def_model} our previous model of transient compartmentalization to the case of 
replicating molecules, which could be involved in uncoupled autocatalytic reactions. We then 
go back to the specific case of competition between ribozymes and parasites originally considered in \cite{Blokhuis2018},
but we go beyond our previous analysis, by including in Sec. \ref{sec: DM}  
deterministic mutations, able to convert ribozymes into parasites. 
The effect of noise in growth dynamics is then covered in Sec. \ref{sec: ng}, which contains, in particular, in 
Sec. \ref{subsec: repmod} a simple model for the replication of a single template by an enzyme, 
and in Sec. \ref{subsec: plnoise} an analysis of the growth noise in a population of replicating polymers.
The latter model is finally used to analyze the effect of noise on the transient compartmentalization dynamics 
introduced in the first section. 

\section{Transient compartmentalization of uncoupled replicating molecules}
\label{sec:def_model}
\subsection{Definition of the model}
Let us formulate an extension of the model of \cite{Blokhuis2018} for general  
replicating molecules in compartments exploiting a common resource.
We start from a pool of molecules, which contains a large number of two types of 
replicating molecules, which we call for simplicity $\ce{A}$ and $\ce{B}$.
Let the fraction of $\ce{A}$ molecules in this pool be $x$.
These molecules then seed a large number of compartments, which is considered to be infinite.
A given compartment will contain $n$ replicating molecules, out of which $m$ will be of $\ce{A}$ type 
and the remaining ones of $\ce{B}$ type. Since this number is small in comparison with the 
number of molecules of the initial pool, $n$ is a random variable drawn from a Poisson distribution of 
parameter $\lambda$, while the number $m$ follows a binomial distribution $B_m(n,x)$.
The resulting probability distribution for seeded compartments is then
\be
P_\lambda(n,m,x)=\text{Poisson}(\lambda,n)B_m(n,x).
\ee

The replicating molecules $\ce{A}$ and $\ce{B}$ are involved in separate autocatalytic reactions, exploiting a common resource $\ce{C}$, yielding the simplified reactions
\bea
\label{autocatalytic}
\ce{A + C } &\ce{<=>}& \ce{2 A + D}, \\
\ce{B + C } &\ce{<=>}& \ce{2 B + F}, 
\eea
where $\ce{D}$ and $\ce{F}$ are product molecules. In principle, we can add other resources and products, as long as we suppose $\ce{C}$ to be limiting. The network may also require catalysts, which for simplicity are not specified in this balance equation. 
Non-replicating molecules and catalysts are assumed to be present in large numbers in the compartments.

After seeding, the numbers of $\ce{A}$ molecules, $m$, and of $\ce{B}$ molecules, $y$, grow exponentially and independently
so that 
\bea
\bar{m}&=&m e^{\alpha T}, \label{mdet}\\
\bar{y}&=&(n-m) e^{\gamma T}, \label{ydet}
\eea
with $T$ the time which marks the end of the exponential growth phase, $\bar{m}$ the number of $\ce{A}$ molecules and 
$\bar{y}$ the number of $\ce{B}$ molecules at time $T$. 
The autocatalytic reactions of Eq.~\eqref{autocatalytic} must eventually saturate at some point 
either because the reaction will run out of fuel molecules or because the waste product molecules 
$\ce{D}$ and $\ce{F}$ poison the reaction. 
For simplicity, let us assume that the growth phase ends when $N = \bar{m} +\bar{y}$, where $N$ is the same
constant for all compartments. Now, the final composition at this end time $T$ is mainly controlled by the ratio $\Lambda=e^{(\gamma - \alpha)T}$. 
Here, we do not describe the saturation 
which could be done more precisely using the notion of carrying capacity \cite{Houchmandzadeh2018}. 
In that case, the growth would be described by logistic equations  
and the carrying capacity would be equal to $N$. 
Note that $N$ can be many times larger than $n$, due to the absence of a division step (which would impose $N \approx 2 n$). This means that a smaller fraction (at least $n/N$) of compartments is enough to carry the functional molecules to the next generation. 
For a dividing cell, on average at least half of its daughter compartments must survive to avoid extinction 
of the population. 

The fraction of $\ce{A}$ molecules at the end of growth phase can be well approximated as 
\be
\bar{x}(n,m)=\frac{\bar{m}}{N} = \frac{m}{n \Lambda - (\Lambda - 1) m}.
\ee
If $\ce{B}$ grows faster, we have $\gamma > \alpha$, and thus $\Lambda>1$, 
which is the regime considered in Ref. \cite{Blokhuis2018}. In Sec \ref{sec: DM}, 
we also consider regimes in which $\gamma < \alpha$. 

We now implement selection at the compartment level. Selection can in general be described by a selection function $0 \geq f(\bar{x}) \geq 1$, which is the fraction of compartments with composition $\bar{x}$ that pass the selection step. In our work, we have assumed 
that the selection function only depends on the final composition $\bar{x}$ of the compartment. 
A natural choice for $f$ is a monotonically increasing function of $\bar{x}$. 
As an example, we will use the sigmoidal function 
\be
f(\bar{x})=0.5\left(1+\tanh \left(\frac{\bar{x}-x_{th}}{x_w}\right)\right), \label{fxpar}
\ee
where $x_{th}$ and $x_w$ are dimensionless parameters, which describe respectively 
a threshold in the composition and the steepness of the function.

The compartments which have passed the selection step are then pooled together, forming a new pool of molecules 
from which future compartments can be seeded. The fraction of $\ce{A}$ molecules, $x'$ of this new ensemble is the average 
of $\bar{x}$ among the selected compartments
\be
x'=\frac{ \langle \bar{x} f(\bar{x}) \rangle}{\langle f(\bar{x} \rangle)},
\ee
which is equivalent to 
\be
x'(\lambda,x)= \frac{\sum_{n,m} \bar{x}(n,m) f(\bar{x}(n,m)) P_{\lambda}(n,x,m)}{\sum_{n,m} f(\bar{x}(n,m)) P_{\lambda}(n,x,m)}. \label{xprime}
\ee
The transient compartmentalization cycle is then repeated, starting with the seeding of new compartments from that pool of composition $x'$. 

Upon repetition of this protocol, the pool composition typically converges to a fixed point $x^*$, 
which is a solution of 
\be
x=x'(\lambda,x).
\ee
Since the variable $x$ undergoes a discrete mapping, 
the fixed point $x^*$ is stable (resp. unstable) if the derivative of $x'(x)$ 
at $x=x^*$ is larger (resp. smaller) than one \cite{Strogatz1994}. 
Therefore, 
the stability of the fixed point $x^{*}$ changes when
\be
\left.\frac{d x'}{dx}\right|_{x=x^*}=1. \label{ddxstab}
\ee

\subsection{Application to ribozyme-parasite dynamics}
The above model has been introduced in Ref.~\cite{Blokhuis2018} to describe replication 
of RNA ribozymes (resp. parasites) in compartments, which play the role of the $\ce{A}$ molecules (resp. $\ce{B}$ molecules). 
In this case, in addition to the replicating molecules, a large amount of $Q\beta$ replication enzymes $n_{Q\beta}$ and activated 
nucleotides $n_u$ (serving as $\ce{C}$ molecules) is supplied in each compartment with the same concentration in each compartment. 
At the end of this growth phase, we have $n_{Q\beta} \approx N = \bar{m} +\bar{y}$, at which point further growth 
is limited by the number of replication enzymes. 
After time $T$, the growth will be linear instead of exponential, but in any case, the system 
composition defined here by the relative fraction of ribozymes, will not change. 
The exact time $T$ could depend on $m,n$, but since in practice $N \gg m,n$ this dependence has a small effect 
on the results of the model 
as we have checked in the Suppl. Mat. of Ref. \cite{Blokhuis2018}.
In Ref \cite{Matsumura2016}, a measurement of the synthesis of a dye molecule by photodetection was used to promote or reject compartments. This selection served as a proxy for more general survival scenarios, acting on the level of the compartment, due to catalytically active RNA. In the following, we consider the specific case of the ribozyme-parasite dynamics of 
Ref.~\cite{Matsumura2016}-\cite{Blokhuis2018}. 

\subsection{Main dynamical regimes}
\begin{figure}
\centering
\includegraphics[scale=0.4]{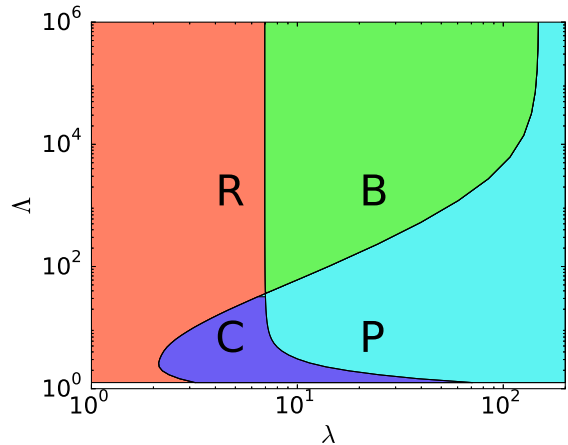} 
\caption{Original phase diagram of the mutation free model, 
taken from Ref\cite{Blokhuis2018}. 
The various phases are pure ribozyme (R), bistable (B), coexistence (C), pure parasite (P).} 
\label{fig: phdiag1}
\end{figure}

Although finding a fixed point $x^{*}$ is generally difficult, our ribozyme-parasite model contains two simple fixed points: $x=0$ and $x=1$. By evaluating the stability of these two fixed points, four regimes can be distinguished, which are shown in the phase diagram in Fig \ref{fig: phdiag1}.
If $x=1$ is stable and $x=0$ unstable, ribozymes are stabilized, and parasites are purged. If $x=0$ is stable and $x=1$ unstable, parasites deterministically invade the pool and purge ribozymes. If both $x=0$ and $x=1$ are unstable, trajectories from either side are attracted to a stable third fixed point $0<x^{*}<1$, leading to stable coexistence between parasites and ribozymes. Finally, if $x=0$ and $x=1$ are stable, their basins of attraction are separated by a third fixed point $0<x^{*}<1$, which is unstable, 
and in this case we have a bistable regime in which the initial composition determines the fate of the system. 

These conclusions can only be drawn provided there are no other fixed points besides ($x=0, x=1, x=x^*$). Extra fixed points come in pairs (one stable, one unstable) and matter only if they are situated within $(0,1)$, in which case a stable coexistence and a bistable phase would be added to the behavior inferred from the other fixed points. For simple monotonically increasing selection functions, we find that extra fixed points are a rare occurrence. Nevertheless, a case where this occurs has been discussed in the Suppl. Mat. of Ref. \cite{Blokhuis2018}.

\subsection{Comparison to experiments}
In addition to predicting the phase diagram associated with the long-time compositions 
reached by this transient compartmentalization dynamics, our theoretical model  
makes also predictions regarding the evolution of the ribozyme fraction as
function of the round number, {\it i.e.} the number of completed cycles 
of compartmentalization. The model correctly reproduces that this fraction 
quickly goes to zero as function of the round number in bulk, less quickly
with compartmentalization and no selection and even less quickly in the case
of compartmentalization with selection. In the latter case,  
a finite fraction can be maintained for an infinite number of rounds 
provided $\lambda$ is sufficiently small, corresponding 
to the coexistence region of the phase diagram.

In order to compare precisely the predictions of the model to the experiments of Ref.~\cite{Matsumura2016},
it is important to know the value of key parameters such as $\Lambda$.
Table \ref{table1} reports the experimental parameters measured in Ref.~\cite{Matsumura2016} 
for the ribozyme and three different parasites. The nucleotide
length, its doubling time ($T_d$), its relative replication rate ($r$) from which we infer 
$\Lambda$ in the final column. The doubling time $T_d$ for the ribozyme is related to the growth rate 
$\alpha$ by $T_d=\ln(2)/\alpha$, and similarly the doubling times of the parasites is $T_d=\ln(2)/\gamma$.

\begin{table}
\begin{center}
\begin{tabular}{|c|c|c|c|c|}
  \hline
  Type  & Length (nt) 2 & $T_{d} (s)$ &  Relative $r$ & $\Lambda$ \\
  \hline
  Ribozyme & 362 & 25.0 &  1.00 & 1\\
  Parasite 1 & 245 & 20.7 &  1.21 & 13\\
  Parasite 2 & 223 & 17.1 &  1.46 & 107\\
  Parasite 3 & 129 & 14.6 &  1.71 & 473 \\
  \hline
\end{tabular}
\end{center}
\caption{Lengths and doubling times for the parasites and ribozyme observed in Ref. \cite{Matsumura2016}, together with their relative aggressivity measured by their relative growth rate $r$, and the corresponding values of $\Lambda$. }  \label{table1}
\end{table}

In the experiment, a typical compartment contains $\lambda$ RNA molecules that can be ribozymes or parasites,
$2.6 \cdot 10^{6}$ molecules of Q$\beta$ replicase, and $1.0 \cdot 10^{10}$ molecules of each NTP. 
Replication takes place by complexation of RNA with Q$\beta$ replicase, which uses NTPs 
to make a complementary copy. This copy is then itself replicated to reproduce the original. 
There is a large amount of nucleotides, so that 
exponential growth of the target RNA proceeds until $N \approx n_{Q \beta}$. 
This large quantity of enzymes also means that in practice, the noise due to fluctuations 
in the number of enzymes should be very small.  
Starting from a single molecule, it takes $n_{D}=\log_{2} n_{Q \beta} =21.4$ doubling times to 
reach this regime. In a parasite-ribozyme mixture, we can estimate $\Lambda$ using the relative $r$:
\be
\Lambda=\frac{2^{n_D}}{2^{n_D / r }}=2^{n_D (1-\frac{1}{r})}.
\label{Lambda}
\ee

\section{A modified model with deterministic mutations}
\label{sec: DM}
In the deterministic model, we assume that a fraction $\mu$ of replicated ribozyme strands mutate into parasites. Thus, the equations describing the evolution of $m$ and $y$ in the growth phase assumes the form
\bea
\dot{m}&=&\alpha m - \mu m = (\alpha - \mu) m \\
\dot{y}&=&\gamma y + \mu m \nonumber,
\eea
which yields for the first equation
\be
\bar{m}=m e^{(\alpha-\mu)T} \label{mbar},
\ee
where $\bar{m}$ is again the number of ribozymes at the end of the growth phase and $m$ the value at the initial time. Now substituting Eq. \eqref{mbar} into the equation for $y$, one finds
\be
\bar{y}=\Big( n-m+\mu m \frac{e^{(\alpha-\gamma-\mu)T}-1}{\alpha - \mu - \gamma}\Big) e^{\gamma T}. \label{ybb}
\ee
The ratio between the number of daughters of one parasite molecule and the number of daughters of a ribozyme molecule 
is now renormalized by the rate $\mu$: $\bar
{\Lambda}=e^{(\gamma + \mu - \alpha) T}=e^{\mu T} \Lambda$, where $\Lambda$ is the relative growth of parasites 
introduced previously in the mutation-free model.

The fraction of ribozymes at the end of the exponential phase is now given by
\be
\label{xb}
\bar{x}(n,m)=\frac{\bar{m}}{N}=\frac{m}{n\bar{\Lambda}-(\bar{\Lambda} -1)(1+\delta) m} , 
\ee
where $\delta=\mu/(\alpha - \mu -\gamma)$. We call $\delta$ the mutation ratio, which is a dimensionless measure of mutation versus relative growth (competition). When $\delta \rightarrow 0$, we recover the mutation-free model, if $|\delta|\gg0$ mutations become dominant.

\begin{figure}
\centering
\includegraphics[scale=0.40]{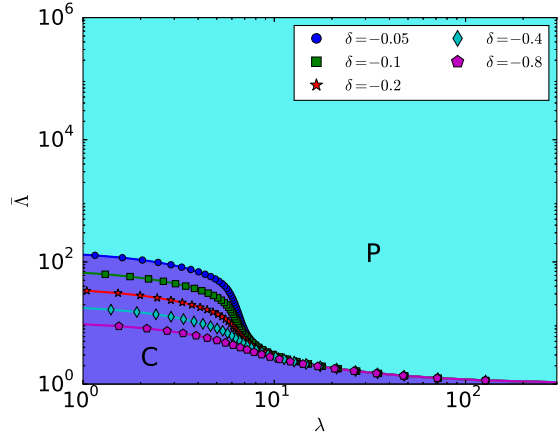} 
\caption{Phase diagram of the model with mutation in the case of prolific parasites. 
The selection function is given in Eq. \eqref{fxpar}. 
Phases are colored for $\delta=-0.05$, other separatrices are plotted for various mutation strengths 
$\delta$. The possible phases are coexistence (C), pure parasite (P). } 
\label{fig: phdiag}
\end{figure}

Selected compartments are then pooled together, and the new average fraction of ribozymes becomes 
$x'(x,\lambda,\delta,\bar{\Lambda})$. Note that for nonzero mutation rate ($\mu>0$), $x' =1$ 
ceases to be a fixed point in this deterministic approach, 
since parasites will always appear at sufficiently long times. 
Therefore, the pure ribozyme (R) phase
is no longer present in the phase diagram of fig. \ref{fig: phdiag}.

The fixed point $x'=0$ however is still present. If this fixed point is stable, we have a pure parasite phase. If it is unstable, 
there is stable coexistence at a fixed composition. If more fixed points appear, multiple 
stable compositions are in principle be possible. 

\subsection{The prolific parasites regime ($\bar{\Lambda} \geq 1$)}

Prolific parasites have a better bulk reproductive success than ribozymes, when 
$\bar{\Lambda} \geq 1$, which is equivalent to $\alpha \leq \mu + \gamma$ and $\delta<0$.
In a mutation-free model, this would imply necessarily a faster growth of parasites ($\alpha<\gamma$), 
but in the present case, we could also allow for slower parasites as compared to ribozymes 
(i.e. $\alpha>\gamma$), provided parasites are aided by a sufficiently high mutation rate $\mu$. 


The phase diagram is evaluated by testing the stability of the fixed point $x'=0$. We find an asymptote behaving like $1/\lambda$ for large $\lambda$, and plateaus for small $\lambda$. The ends of these plateaus locate in the limit $\delta \rightarrow 0$ at the position of the vertical line separating the ribozyme and bistable phase in the original phase diagram.

Let us first derive the right asymptote in the $\lambda \gg 1$ limit. In this limit, we evaluate $x'$ by considering  
compartments of size $\lambda$
\be
x'=\frac{\lambda x \bar{x} f(\bar{x}) }{(1-x) f(0) + \lambda x f(\bar{x})}.
\ee
The fixed point stability condition $\left. d x'/ dx \right|_{x=0}=1$ leads to
\be
\left.\frac{d x'}{dx}\right|_{x=0}=\frac{\lambda \bar{x} f(\bar{x})}{f(0)}.
\ee
Upon substituting Eq. \eqref{xb} evaluated at $m=1, n=\lambda$ and approximating $f(\bar{x}) \approx f(0) + f'(0) \bar{x}$, (for $\lambda \gg 1, \bar{x} \ll 1$) we find a quadratic equation for $\bar{\Lambda}$, whose only physical solution ($\bar{\Lambda}\geq 1$) is 
\be
\bar{\Lambda}=\frac{ \lambda - 2 \delta - 2 + \sqrt{\lambda \left(4 \frac{f'(0)}{f(0)} + \lambda \right)  }}{2 (\lambda - \delta -1)}.
\ee
Since we consider monotonically increasing selection functions, $f'(0)>0$. For $\lambda \gg - \delta$, we find
\be
\bar{\Lambda}=1+\frac{f'(0)}{f(0) (\lambda - \delta - 1)} \approx 1 +\frac{f'(0)}{f(0)  \lambda},
\ee
which is the same expression as the one found in the mutation-free phase diagram \cite{Blokhuis2018}.
This explains why there is a single asymptote as $\mu$ is varied in the $\lambda \gg 1$ limit.

The plateaus extend to very low values of $\lambda$. We can find their location by considering only compartments of size $n=1$. 
In that case, the final compositions can be $\bar{x}(1,0)=0$ or 
\be
\bar{x}(1,1)=\frac{1}{1+\delta -\delta \bar{\Lambda}}. \label{xb1}
\ee
We then have for the composition recursion
\be
x'= \frac{ x \bar{x} f(\bar{x}) }{ (1-x)  f(0) +  x  f(\bar{x})}. \label{xx1}
\ee
Evaluating the derivative of $x'(x)$, we find
\be
\frac{\bar{x}f(\bar{x})}{f(0)}=1.
\ee
Substituting \eqref{xb1}, we find that the location of plateaus obeys the implicit equation 
\be
\bar{\Lambda}=1+ \frac{f(0) - f(\bar{x})}{f(0) \delta}. \label{paq}
\ee

\subsection{The prolific ribozymes regime ($\bar{\Lambda}\leq 1$)}

We now consider the opposite case where parasites are less prolific than ribozymes. This 
means $\alpha \geq \mu + \gamma$ and is equivalent to 
$\bar{\Lambda}\leq 1, \delta>0$. This implies that $\alpha>\gamma$ (less aggressive parasites) and 
is reminiscent of a quasipecies scenario in which a fit ribozyme successfully outcompetes its parasites in bulk \cite{Eigen1971}.
Since this can already happen in the absence of selection, we consider here the case where there is no selection, {\it i.e.} 
 $f(\bar{x})=1$.

\begin{figure}
\centering
\includegraphics[scale=0.5]{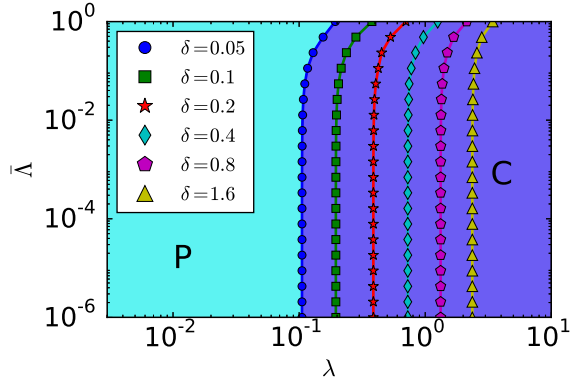} 
\caption{Phase diagram in absence of selection function for prolific ribozymes ($\bar{\Lambda} \leq 1$). Phases are colored for $\delta=0.05$, separatrices are plotted for various mutation strengths $\delta$. C: coexistence, P: pure parasite. } 
\label{fig: phdiag4}
\end{figure} 
To analyze this regime, we again assess the fixed point stability of $x'=0$. We locate numerically the separatrix as shown in 
Fig \ref{fig: phdiag4}. We obtain separatrices that for $\bar{\Lambda} \rightarrow 0$ tend to a fixed value of $\lambda$.

Let us start by observing that when $\bar{\Lambda}\rightarrow 0$, there are only two final compartment compositions for nonempty compartments: $\bar{x}(n,0)=0$ or $\bar{x}(n,m)=1/(1+\delta)$ for $m>0$. We can now distinguish between three initial compartment compositions: (i) only parasites,  (ii) no parasites, no ribozymes, and (iii) containing at least one ribozyme. Their associated seeding probabilities are:
\bea
p_{\text{para}}&=&\sum_{n=1}^{\infty}\frac{(1-x)^{n}\lambda^{n}}{n!} e^{-\lambda}=(e^{\lambda (1-x)}-1) e^{-\lambda} \nonumber  \\
p_{\text{zero}}&=&e^{-\lambda} \\
p_{\text{ribo}}&=&1-p_{\text{para}}-p_{\text{zero}}=1-e^{-\lambda x} \nonumber
\eea

In that case, we can write the composition recursion equation as 
\be
x'= \frac{1}{1+\delta} \frac{p_{\text{ribo}}}{p_{\text{para}} + p_{\text{ribo}}},
\ee
The condition $\left. d x'/dx \right|_{x=0}=1$ yields the expression
\be
\lambda =(1+\delta)(1-e^{-\lambda}), \label{vas}
\ee
for the asymptote. For $\lambda \ll 1$, 
we obtain using \eqref{vas} 
\be
\lambda= \frac{2 \delta}{1+\delta},
\ee
which agrees very well with Fig \ref{fig: phdiag4}.

Notice that here the coexistence phase is located to the right of the asymptotes, and the parasite phase to the left, whereas in Fig \ref{fig: phdiag} it is the other way around. An intuitive way to understand this is to consider the limit $\lambda \rightarrow 0$. In this limit, nonempty compartments start with either a parasite or a ribozyme. The former will grow to a fully parasitic compartment, whereas the latter will contain ribozymes plus some parasites acquired by mutations. Therefore, at low $\lambda$, the ribozyme's capacity to outgrow parasites (competition) cannot be exploited, leading to ribozyme extinction. It is only when ribozymes and parasites are seeded together that the differential growth rate becomes important, which becomes increasingly likely for higher $\lambda$. The phase boundaries in Fig. \ref{fig: phdiag4} mark the point where enough compartments engage in competition to allow for ribozyme survival. The mutation strength $\delta$ compares mutation rate to competition. When $\delta \rightarrow 0$, there is enough competition to ensure coexistence for all $\lambda$. 

\subsection{Error catastrophe}

An error catastrophe corresponds to a situation where the accumulation of replication errors eventually causes the disappearance of ribozymes. Since there are only a parasite (P) and a coexistence phase (C) in the model with mutations, the error catastrophe means that the coexistence region shrinks at the benefit of the parasite phase as the mutation rate increases. 
One sees this effect in Fig. \ref{fig: phdiag}, which corresponds to the prolific parasites regime ($\bar{\Lambda} \geq 1$) discussed above. 
In this figure, we see a larger coexistence region in the small $\lambda$ region, because there the compartmentalization is efficient
 to purge parasites. As the mutation rate increases however, this region shrinks because the compartmentalization fails to purge the 
more numerous parasites.

\begin{figure}[!tbp]
  \centering
  \begin{minipage}[b]{0.49\textwidth}
    \includegraphics[width=\textwidth]{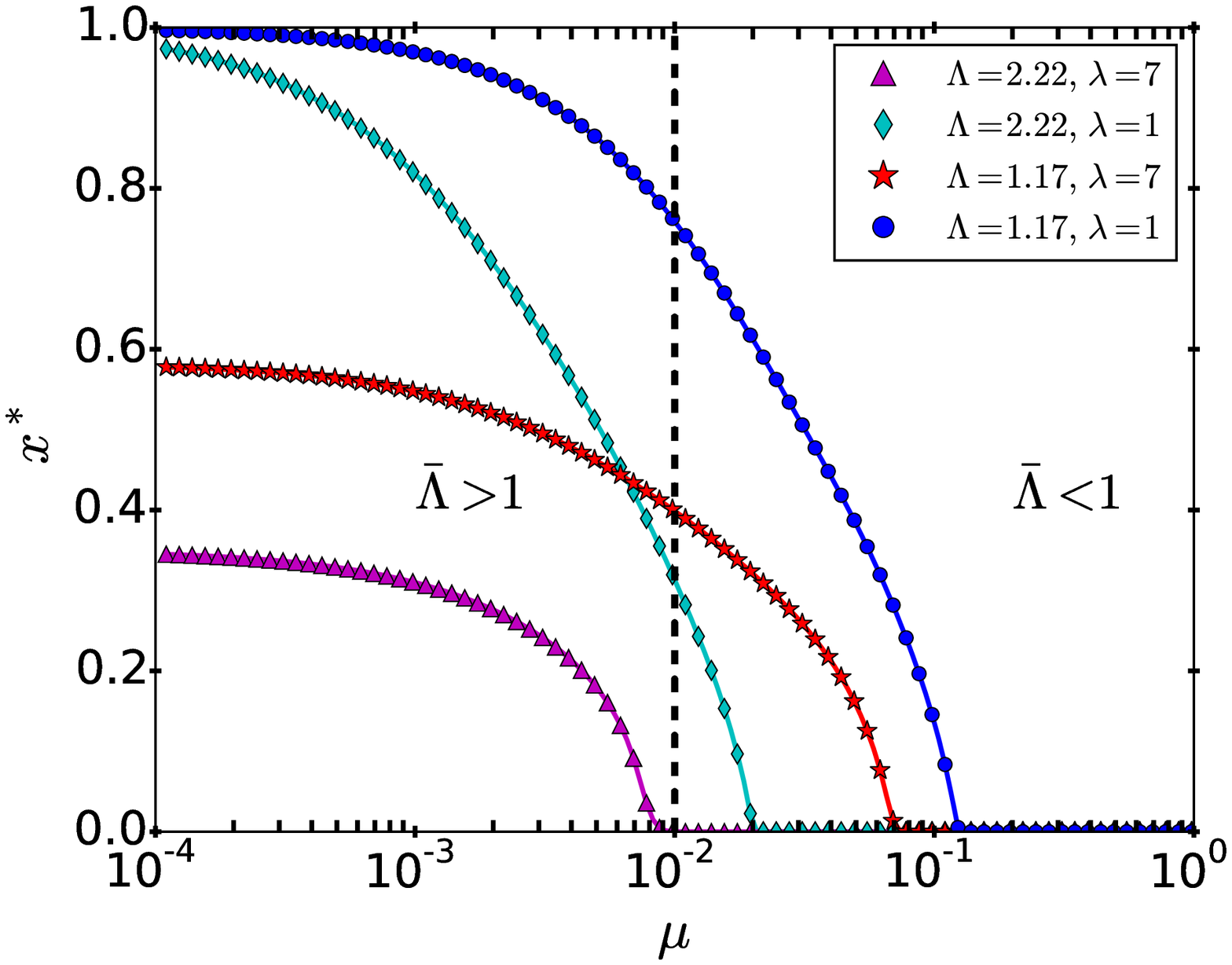}
    \caption{Steady state composition $x^*$ as function of $\mu$, $\alpha=0.99, \gamma=1.0$. Critical rates $\mu^*$ corresponds to separation between P and C phases in Fig. \ref{fig: phasediagF}. }
    \label{fig: errcatF}
  \end{minipage}
  \hfill
  \begin{minipage}[b]{0.50\textwidth}
    \includegraphics[width=\textwidth]{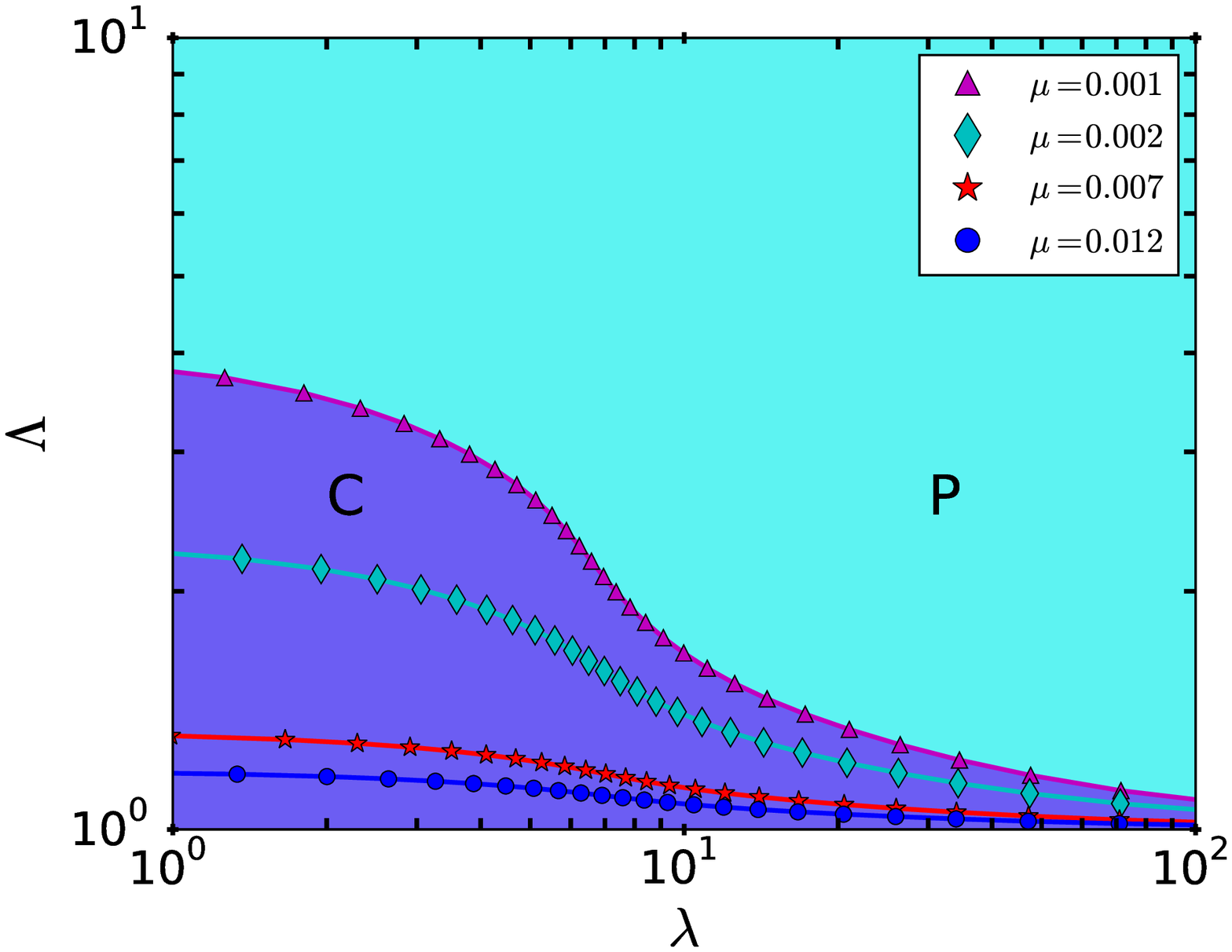}
    \caption{Phase diagram, drawn for f$\alpha=0.99, \gamma=1.0$. Separatrices are drawn for $\mu$ values close to $\mu^*$ in Fig. \ref{fig: errcatF}, corresponding to an error catastrophe.}
    \label{fig: phasediagF}
  \end{minipage}
\end{figure}

\begin{figure}[!tbp]
  \centering
  \begin{minipage}[b]{0.48\textwidth}
    \includegraphics[width=\textwidth]{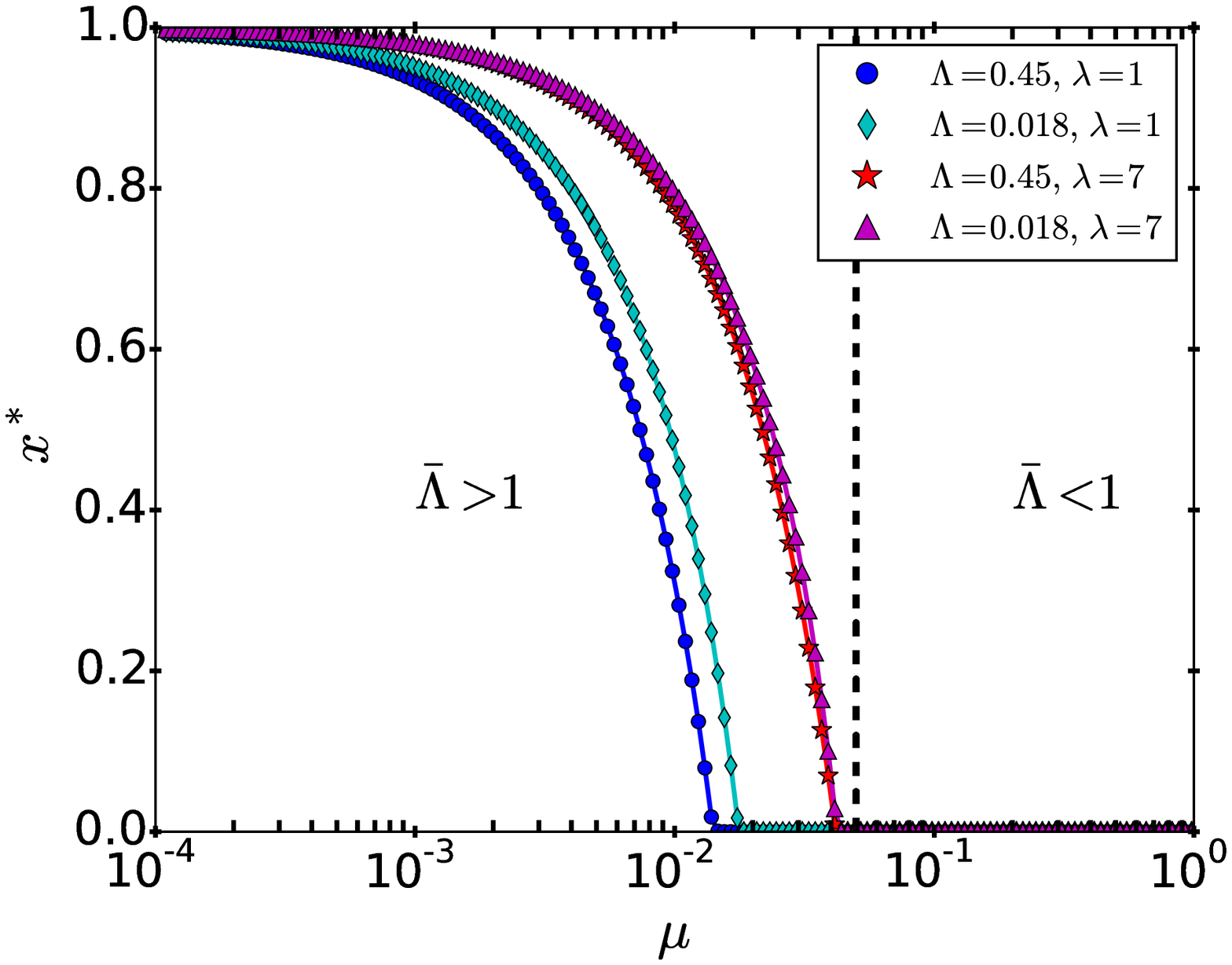}
    \caption{Steady state composition $x^*$ as function of $\mu$, $\alpha=1.0, \gamma=0.95$, in absence of selection ($f=1.0$). Critical rates $\mu^*$ corresponds to separation between P and C phases in Fig. \ref{fig: phasediagf1} .}
    \label{fig: errcatf1}
  \end{minipage}
  \hfill
  \begin{minipage}[b]{0.51\textwidth}
    \includegraphics[width=\textwidth]{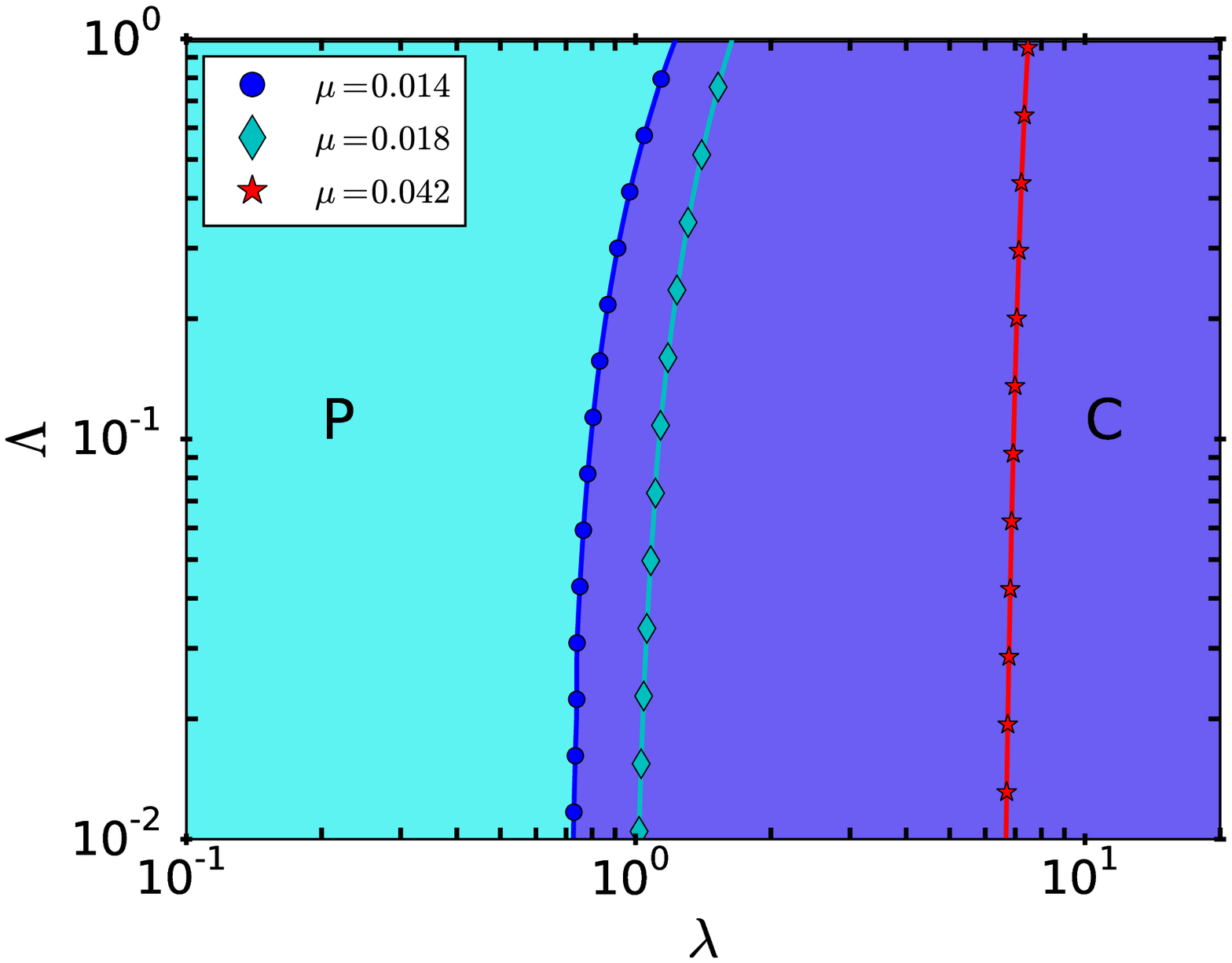}
    \caption{Phase diagram in absence of selection function ($f=1.0$), drawn for $\alpha=1.0, \gamma=0.95$. Separatrices are drawn for $\mu$ values close to $\mu^*$ in Fig. \ref{fig: errcatf1}.}
    \label{fig: phasediagf1}
  \end{minipage}
\end{figure}

In Fig. \ref{fig: errcatF}, a particular example is provided where $\alpha$ and $\gamma$ are fixed, such that $\bar{\Lambda}$ is fixed, and $\mu$ is varied. Since competition is fixed, we have $\mu \propto \delta$. The resulting steady-state value $x=x^*$ then decreases monotonically with $\mu$, and reaches $x=0$ when crossing the phase boundary in Fig \ref{fig: phasediagF}. For small values of $\lambda$, this boundary corresponds to the plateau region, for larger values, this corresponds to the $1/\lambda$ asymptote. As can be seen in Fig \ref{fig: errcatF}, coexistence is stable for much higher values of the mutation rate $\mu$ when the compartment size $\lambda$ is small. This means that compartmentalization with selection leads to a relaxed error threshold with respect to the bulk.

The error catastrophe was also studied in the absence of selection and was shown to be in the prolific ribozymes regime ($\bar{\Lambda} \leq 1$). In Fig. \ref{fig: errcatf1}, an example of this case is shown, and there too, we see that the steady-state value of the ribozyme fraction $x^*$ decreases as $\mu$ is increased, until it reaches the phase boundary in Fig \ref{fig: phasediagf1}. In contrast to Fig. \ref{fig: errcatF}, where the error threshold decreases as the size of compartments increases, the trend is just the opposite in Fig. \ref{fig: errcatf1}, which is expected since the role of ribozymes and parasites are exchanged here as compared to the prolific parasites regime.

In the prolific parasites regime, $\bar{\Lambda} \leq 1$ with selection, it is interesting to recast the error threshold as a constraint on the length of a polymer to be copied accurately, as done in the original formulation of the error threshold \citep{Eigen1971}. Let us introduce the error rate per nucleotide, $\epsilon$. Then, for a sequence of length $L$, we have $\alpha - \mu = \alpha (1- \epsilon)^L$. Since $\epsilon \ll 1$, it follows from this that $\mu = \alpha \epsilon L$.   
When $\alpha \simeq \gamma$, we have $\ln \bar{\Lambda}= \alpha \epsilon L T$.
Using Eq. \eqref{paq}, we find that the condition to copy the polymer accurately is 
\be
L \le \frac{\ln(s)}{\epsilon \alpha T},
\ee
where $s=f(\bar{x})/f(0)$ and $\alpha T / \ln 2$ is the number of generations.   This criterion has a form similar to the
 original error threshold \cite{Eigen1971}, namely
\be
L \le \frac{\ln(s')}{\epsilon},
\ee
where $s'=\alpha/\gamma$ represents the selective superiority of the ribozyme. 
In our model, the equivalent of $s'$ is $s$ which characterizes the compartment selection.

\section{Noise in growth}
\label{sec: ng}

For deterministic growth, given by Eqs. \eqref{mdet}-\eqref{ydet}, 
fluctuations in the growth rates, denoted $\alpha$ for $\ce{A}$ molecules and $\gamma$ for the $\ce{B}$ molecules, 
have been neglected. 
In order to estimate the magnitude and effect of fluctuations in the growth rates, we introduce in the next section a model for noisy replication. 
In particular, we consider a replication enzyme that stochastically binds to a strand, followed by the stochastic incorporation of $L$ monomers. The model can either have i) a single rate-limiting step or ii) L rate-limiting steps. Case i) corresponds to simple autocatalytic reactions, or the rate-limiting binding of a replication enzyme. Case ii) corresponds to the rate-limiting polymerization of a polymer of length $L$, via a multistep replication process. For $L=1$, all these descriptions become equivalent.

Importantly, this model assumes that the replicase, once bound, stays 
active until completion of the copy of the template. The possibility that the replicase 
falls off the template before completion of the copy is neglected. 
Similarly, any effects associated with the interaction 
of multiple replicases on the same template are neglected. 
In fact, when the replicase falls off of its template, the copying process is aborted and the shorter 
chain which has been produced in this way becomes a parasite. We can therefore 
describe such a process as a mutation using the framework of the previous section.
To separate the effects due to mutations and noise clearly, we disregard from now on the possibility of mutations, 
and we focus in the following on the description of the noise associated with replication. Such a noise 
can stabilize the ribozyme phase at the expense of coexistence, and the coexistence phase at the expense of the parasite phase. The noise of replication becomes very small when the rate-limiting step is nucleotide incorporation, in which case one can use a deterministic approach. In case of a single rate limiting step, we obtain giant fluctuations.

\subsection{A minimal model for the replication process}
\label{subsec: repmod}

The replication of a polymer strand $\ce{A}$ by a replicase $\ce{E}$ can be considered to proceed through two stages. In the first stage, a strand $\ce{A}$ binds to a replicase $\ce{E}$, to form a complex $\ce{X_0}$
\be
\ce{A + E} \xrightarrow{\kappa_C} \ce{X_0}  \label{AE},
\ee
with the rate $\kappa_C$. 

Subsequently, activated nucleotides $X$ are incorporated in a stepwise fashion to the complementary strand. A complex of $E$ and $A$ with a complementary strand of length $n$ will be denoted by $X_n$, and the strand grows until the final length $L$ is achieved, such that
\bea
\ce{X_n + X} &\xrightarrow{\kappa}& \ce{X_{n+1}}, \ \  0 \leq n \leq L-2 \\ \label{XN}
\ce{X_{L-1} + X} &\xrightarrow{\kappa}& \ce{2 A},
\eea
where for simplicity we have assumed the same rate $\kappa$ for both reactions.
Let us denote by $t$ the total time to yield $\ce{2 A}$ from $\ce{A}$, which is the sum of the time associated with the step of complex formation, $t_C$ and with the step of $L$ nucleotide incorporations $t_L$. We thus have 
\be
t=t_C + t_L ,
\ee
with $t_L=\sum^L_{i=0} t_i$ and $t_i$ the time for adding one monomer, which we assumed is distributed according to
\be
f(t_i)=\kappa e^{-\kappa t_i}.
\ee
For simplicity, we choose a single value $\kappa$ for all monomer additions. 
The time for the formation of the complex, $t_C$ is similarly distributed according to 
\be
f(t_C)=\kappa_C e^{-\kappa_C t_C}, \label{tcomplex}
\ee
where $\kappa_C=1/\langle t_C \rangle$.

Let us denote the moment generating function of $t_C$ by $M_C(s)$ and similarly for $t_L$ by $M_L(s)$ with : 
\bea
M_{C}(s)&=&\int^{\infty}_0 dt_C \exp( - s t_C) f(t_C) \nonumber \\
&=&\frac{\kappa_C}{s+\kappa_C} , \label{MD}\\ 
M_{L}(s)&=&\int^{\infty}_0 dt_L \exp(- s t_L) f(t_L) \nonumber \\
&=& \prod^{L}_{i=1} \Big[\int^{\infty}_0 dt_i \exp(- s t_i) f(t_i) \Big] \nonumber \\
&=&\left(\frac{\kappa}{s+\kappa}\right)^{L}. \label{ML}
\eea

From $M_L$ one obtains the distribution of replication time $f(t_L)$ by performing an inverse Laplace transform:
\be
f(t_L)=\mathcal{L}^{-1} \left[M_L(s)\right]= \frac{\kappa^L t_L^{L-1} e^{-\kappa t_L}}{\Gamma(L)}, \label{pgam}
\ee
where $\mathcal{L}^{-1}$ represents the inverse Laplace transform. This equation shows that the replication time distribution of one strand of length $L$ follows a Gamma distribution \cite{Floyd2010}. For $L=1$, Eq. \eqref{pgam} becomes a simple exponential distribution, which is a memoryless distribution. This distribution describes any process with a single rate-limiting step, such as simple autocatalysis or the binding of the replicase. 

For $L>1$, this distribution has memory and the growth in the number of RNA strands can no longer be described as a simple Markov process. 
Note that the Gamma distribution is peaked around the mean value of $t_L$, namely 
$L/\kappa$ for $L \gg 1$. In this limit, the replication time has very small fluctuations. 
This feature has recently been exploited to construct a single-molecule clock, 
in which the dissociation of a molecular complex occurs after a well-controlled 
replication time\cite{Johnson-Buck2017}.

\subsection{Coefficient of variation of the replication time}
\label{subsec: RLG}
\begin{figure}
\centering
\includegraphics[scale=0.50]{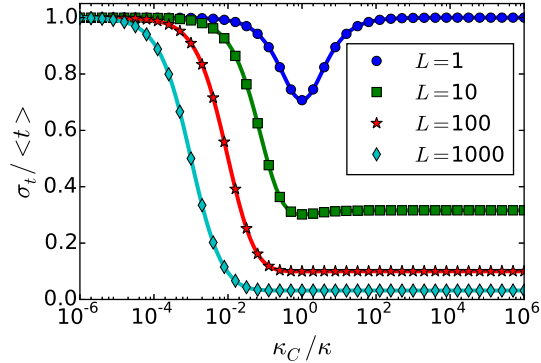} 
\caption{Waiting time variability $\sigma_t/\langle t \rangle$ for various polymer lengths $L$, 
as a function of the ratio of typical times for replication and complex formation} 
\label{fig: Lplot}
\end{figure}

Let us now study the coefficient of variation of the full time $t$. For the simple replication model, this includes 
the diffusion of the replicase and the replication step. 
The generating function of $t$ is clearly $M(s)=M_D(s) M_L(s)$. Thus,  
the cumulant-generating function defined as $K(s)=\ln M(s)$, 
yields the two moments of the distribution of $t$,  namely
the mean $\langle t \rangle$ and the variance $\sigma_t^2$. We have 
\bea
\langle t \rangle=\langle t_{C} \rangle+ \langle t_{L} \rangle = \frac{1}{\kappa_C} + \frac{L}{\kappa}, \\
\sigma_t^2=\sigma^2_{C}+\sigma^2_{L} = \frac{1}{\kappa^2_C} + \frac{L}{\kappa^2}.
\eea
Thus the coefficient of variation of the replication time, namely $\sigma_t/ \langle t \rangle$ is given by
\be
\frac{\sigma_t}{\langle t \rangle}=\frac{\sqrt{\frac{1}{\kappa^2_C} + \frac{L}{\kappa^2}}}{\frac{1}{\kappa_C} + \frac{L}{\kappa}}. \label{coefvar}
\ee
Fig \ref{fig: Lplot} shows this quantity as function of the length $L$ and of the ratio of the rates ($\kappa_C/\kappa$). 

There are two regimes: on one hand, when $L/\kappa \gg 1/\kappa_C$, the time taken by the replication step dominates over the time for 
the replicase to diffuse to its target. If in addition $\sigma^2_L \gg\sigma^2_C$,
 the coefficient of variation of the time $t$ scales as $1/\sqrt{L}$ and therefore becomes 
very small for long strands. This power-law regime is indeed visible as plateaus in Fig \ref{fig: Lplot} and we will refer to this as the replication-limited regime.

On the other hand, when $1/\kappa_C \gg L/\kappa$, the time to form a complex between the replicase and its template dominates over the replication time. This regime has a large coefficient of variation since $\sigma_t \simeq \langle t \rangle$
as also seen in Fig \ref{fig: Lplot}. In this regime, the replication time is governed by the simple exponential distribution of Eq. \eqref{tcomplex}. A simple autocatalytic reaction is governed by such a distribution, which is also equivalent to a replication-limited situation with $L=1$. We will refer to this behavior as the diffusion-limited regime.

\subsection{Phylogenetic noise due to asynchronous growth}
\label{subsec: genrep}

\begin{figure}
\centering
\includegraphics[scale=0.60]{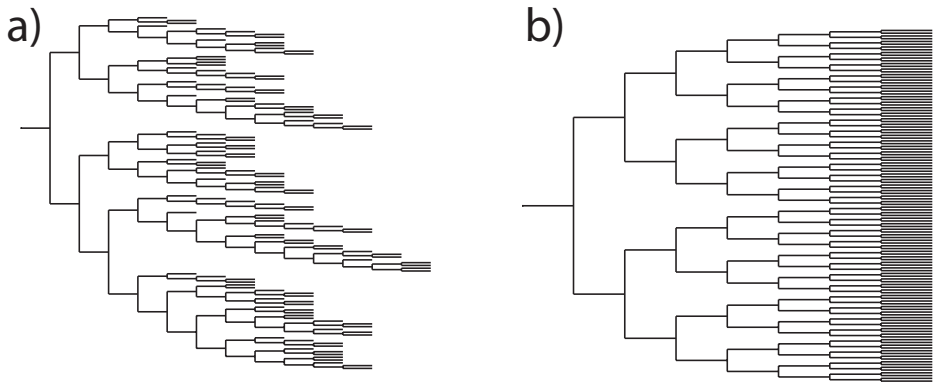} 
\caption{Phylogenetic trees, generations representation. a) diffusion-limited regime. b) replication limited regime. The simulation ends when the population size has reached $128$. The horizontal axis corresponds to the generation number.} 
\label{fig: Growth}
\end{figure} 

In Fig. \ref{fig: Growth}, phylogenetic trees are drawn for diffusion-limited and replication-limited growth. In both cases, growth starts from a single parent strand and descendants are depicted as function of their generation. 

In this representation, the differences in the two growth regimes become very clear. In the replication-limited regime, generations are synchronized: lineages spread over the same numbers of generations. This happens because noise in replication time is small with respect to the typical replication time:  $\sigma_t / \langle t \rangle \ll 1$  (see Eq. \eqref{coefvar}). Replication slowly desynchronizes by the accumulation of noise over multiple generations. For two independent strains, generations become desynchronized after about $\sqrt{L}$ generations. 

In contrast, in the diffusion-limited regime, fluctuations are of the order of the replication time: $\sigma_t / \langle t \rangle = 1$. In this memoryless case, each species is equally likely to perform the next replication event, yielding a desynchronized growth behavior with large gaps in the phylogenetic tree.

These figures have been obtained by simulating the growth of a replicating mixture starting from a single strand. The simulation follows $k$ RNA-enzyme complexes, and for each the variable $n_k$ measures the length of the growing complementary strand. For every nucleotide incorporation event, a strand $i$ is chosen with probability $1/k$, after which its number of nucleotides is updated from $n_i$ to $n_i+1$. When $n_i+1=L$, we set $n_i=0$, we update $k$ to $k+1$, and then we introduce an extra strand variable $n_{k+1}$ for the new strand. 
Both the replication-limited regime and the diffusion-limited regime can be modeled using this simulation. In the latter case, we choose $L=1$, which corresponds to exponentially distributed replication times as in Eq. \eqref{tcomplex}. This also describes the case of simple autocatalysis.

\subsection{Noise in population size due to growth}
\label{subsec: plnoise}
In sec \ref{subsec: RLG}, we have analyzed the noise associated with the replication of a single strand. Ultimately, we wish to 
quantify the compositional variation of the final population. In order to do so, we turn to the theory of branching processes 
with variable lifetimes taken randomly from a fixed distribution \cite{Karlin1975}. As explained in \ref{sec: A}, 
this framework describes theoretically a population that grows exponentially starting from a single individual. 
In our molecular system, this single individual plays the role of the single molecule present in the initial 
condition before the replication starts; while the distribution of the lifetimes  
is the replication time distribution $f(t_L)$ obtained in Eq \eqref{pgam}.

	For $t_L \gg L/\kappa$, we find that the average population (starting from a single individual) $\mu^{(1)}$ scales as $\mu^{(1)} (t) = \mu^* e^{\alpha t}$,
with a growth rate $\alpha \simeq \kappa \ln(2)/L$.
The coefficient of variation of the population size $\sigma^{(1)}/\mu^{(1)}$ is 
\be
\frac{\sigma^{(1)}}{\mu^{(1)}} \approx \frac{\sqrt{2} \ln(2) }{\sqrt{L}}. \label{sqL}
\ee
 
\begin{figure}
\centering
\includegraphics[scale=0.60]{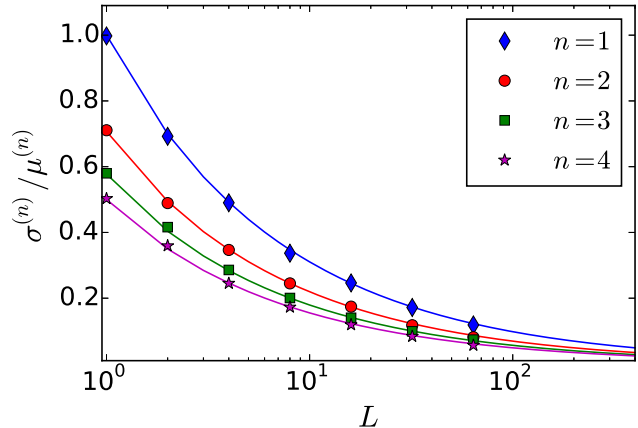} 
\caption{Coefficient of variation of the population
size $N$ as function of the initial population size $n$. 
The results have been averaged over 2000 runs. 
The solid lines represent the theoretical prediction: $1/\sqrt{nL}$. }
\label{fig: coeffit}
\end{figure}

The renewal theory on which these results are based, can be generalized to the case that there are $n$ individuals 
in the initial condition as shown in \ref{sec: B}. The full solution is found by treating the $n$ initial molecules as $n$ independent subpopulations, which all 
start at size $1$ and follow the branching process described above 
and in \ref{sec: A}. In that case, each subpopulation now has a mean $\mu^{(1)}= \mu^{(n)} / n$ and a 
standard deviation $\sigma^{(1)} \approx \mu^{(1)}/ \sqrt{L}$. This then allows to write
\be
\label{sqL_fin}
\sigma^{(n)} \approx \sqrt{ n} \sigma^{(1)} = \frac{\mu^{(1)}}{\sqrt{n L}}.
\ee
We show in Fig. \ref{fig: coeffit} that the corresponding 
coefficient of variation, $\sigma^{(n)}/\mu^{(n)}$, agrees well with simulations of the branching
 process. The 2000 simulation runs were stopped after a time $t^*$ such that 
$\langle N(t^*) \rangle \simeq 5000$.

\subsection{Giant fluctuations in logistic growth of competing species}
\label{subsec: competition}

The problem of two species competing for the same resources has been studied 
in the literature and offers a complementary perspective on the role of noise
in a growing population, which has been studied in the previous section. 
Let us consider two such species, which typically start with a few individuals and then grow according to
 logistic noise. As shown in Ref. \cite{Houchmandzadeh2018},
when the carrying capacity is reached, the number of each species 
is subject to giant fluctuations (the coefficient of variation is of the order of unity) 
when the two species have similar growth rates, leading to considerable compositional variability.
In the terminology introduced in previous section,  this model  
applies to the diffusion-limited regime ($L \to 1$, simple autocatalysis), where 
a Markov description of the population dynamics is applicable.

Keeping the notations of the first section,   
we denote by $n$ the initial number of molecules, 
which splits into $m$ ribozymes (or $\ce{A}$ autocatalysts) and $y$ parasites (or $\ce{B}$ autocatalysts), 
and by $N$ the final number of molecules in the compartment.
In the neutral case ($\alpha=\gamma$), the moments of the number of ribozymes $\bar{m}$
are found to be \cite{Houchmandzadeh2018} : 
\bea
\langle \bar{m} \rangle &=& N \frac{m}{n}, \\
\sigma_{\bar{m}} &=& \ \sqrt{\frac{my}{n^2}  \frac{N (N - n)}{n +1}},
\eea
with again $y=n-m$. Since $N$ remains fixed, $\sigma_{\bar{x}}=\sigma_{\bar{m}}/N$. This means that 
\be
\label{logisticgrowth}
\sigma_{\bar{x}}\approx \frac{1}{n}\sqrt{\frac{my}{n}}
\ee
for $N \gg n$, which means that the variability in the final composition $\bar{x}$ depends primarily on the number of individuals in the initial condition. 
Let us denote $s=\alpha/\gamma -1 \ll 1$, with $s\ll1$ and 
$\rho=\ln(N/n)$. In Ref. \cite{Houchmandzadeh2018}, it was shown that 
\be
\frac{\sigma_{\bar{m}}}{ \langle \bar{m} \rangle} = \sqrt{ \frac{y}{m (n+1)}} 
\left(1- \frac{\rho s n (m+1)}{(n+1)(n+2)}\right) \label{sigB}
\ee 
In general, the dynamics of the composition has a large variability for: 
(i) small compartments ($n \sim O(1)$), (ii) mixed compartments ($m,y>0$),
and for $m \approx y$, (iii) comparable growth rates ($s \rightarrow 0$).

Such a coefficient of variation is asymptotically constant on long times 
and the constant only depends on the initial number of molecules. 
A similar scaling for the coefficient of variation 
holds in a number of other physical situations, such as for the fluctuations in the number 
of protein filament formed in small volumes \cite{Michaels2018}.

\subsection{Noise for co-encapsulated growing populations}
\label{subsec: competition2}

Let us now apply the results of the section \ref{subsec: plnoise} 
to analyze the effect of the growth noise on our transient compartmentalization dynamics.
Let us assume that the length of the ribozymes is $L_\alpha$ and that of the 
parasites $L_\gamma$. For experimental values of these parameters 
we refer the reader to Table \ref{table1}. In section \ref{sec:def_model}, we have 
defined $m, y$ to be the initial number of ribozymes and parasites 
and $\bar{m}, \bar{y}$ to be the final mean number of ribozymes and parasites 
at the end of the growth phase in a given compartment.
Using Eqs. \eqref{sqL}-\eqref{sqL_fin}, we obtain 
\bea
\frac{\sigma_{\bar{m}}}{\langle \bar{m} \rangle} & \simeq & \frac{1}{\sqrt{L_\alpha m}}, \nonumber \\
\frac{\sigma_{\bar{y}}}{\langle \bar{y} \rangle} & \simeq & \frac{1}{\sqrt{L_\gamma y }}. 
\eea

Since the ribozyme fraction $\bar{x}$ at the end of the exponential phase is given by $\bar{x}(n,m)=\bar{m}/N$ and $N \simeq n_{Q \beta}$, the standard deviation of the final composition $\bar{x}(n,m)$ takes 
the following form :
\bea
\sigma_{\bar{x}} &=& \sqrt{\left( \frac{\partial \bar{x}}{\partial \bar{m}} \right)^{2} \sigma^2_{\bar{m}} +
 \left( \frac{\partial \bar{x}}{\partial \bar{y}} \right)^{2} \sigma^2_{\bar{y}}}  , \nonumber \\
&\simeq&\sqrt{ \left( \frac{\bar{y}}{N^2} \right)^{2} \frac{\bar{m}^2}{m L_\alpha} + \left( \frac{-\bar{m}}{N^2} \right)^{2} \frac{\bar{y}^2}{y L_\gamma}} , \label{sigXX} \\
&\simeq &\bar{x} (1-\bar{x}) \sqrt{\left( \frac{1}{m L_\alpha}+\frac{1}{y L_\gamma} \right)} , \nonumber
\eea
where we have used Eq.~\eqref{sqL_fin} with $\mu_{\bar{m}}=\bar{m}, \mu_{\bar{y}}=\bar{y}$. 
The factor $\bar{x}(1-\bar{x})$ is 
largest for $\bar{x}=1/2$ and vanishes for pure parasite and pure ribozyme compartments,
which means that compositional variability due to stochastic growth can be neglected 
when $\Lambda \gg 1$ or $\Lambda \ll 1$.
Note that if we choose $\alpha=\gamma$ (and thus $\bar{x}=m/n$),  and $L_{\alpha}=L_{\gamma}=1$, Eq. \eqref{sigXX} becomes
\be
\sigma_{\bar{x}} \simeq \frac{1}{n}\sqrt{\frac{m y}{n}}
\ee
which is consistent with Eq. \eqref{logisticgrowth} which was found using a different formalism\cite{Houchmandzadeh2018}.

Eqs. \eqref{sigB} and \eqref{sigXX} point to an interesting trade-off : the synchronization of growth rates comes at the cost of greater compositional variability. To have a stable coexistence, growth rates should not diverge too much. However, this also implies giant fluctuations in final composition. In the presence of strong selection, noise in growth will generate many non-sustainable compositions, lowering the overall survival of compartments \cite{Grey1995}.

This reduction in survival is expected to be particularly detrimental if a compartment splits into two daughter compartments \cite{Grey1995}. To prevent extinction, at least half of the daughters should, on average, survive. This puts a constraint on more advanced selection mechanisms, such as the Stochastic corrector. In transient compartmentalization, only a much smaller fraction 
of the order of $\lambda/N$ compartments need to survive. 
The RNA experiments are indeed in this regime since $N \approx O(10^6), \lambda = O(1)$).

By having multiple rate-limiting steps ($L > 1$), compositional variability due to noise in growth is reduced. In this sense, polymerization on a template as considered here is inherently functional: the noise suppression it permits can increase the average survival rate of compartments. Noise suppression also increases evolvability, by giving the system access to more efficient mechanisms of heritability.

\begin{figure}
\centering
\includegraphics[scale=0.50]{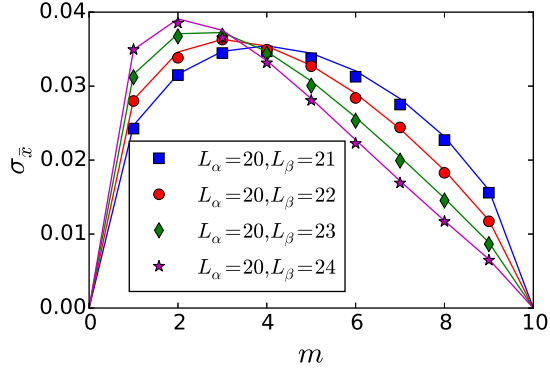} 
\caption{Standard deviation of the ribozyme fraction, $\sigma_{\bar{x}}$, 
as predicted from simulations (symbols), and compared with predictions from 
Eq.~\eqref{sigXX} (solid lines).
For each initial composition $(m,n)$, 10000 simulations were performed until a  
time $t^*$ such that $\langle N(t^*) \rangle \simeq 5000$ and 
by choosing $\alpha/\gamma=L_{\gamma}/L_{\alpha}$.}
\label{fig: sigX}
\end{figure} 

Using the parameters of Table \ref{table1} and \eqref{sqL}, 
we can quantify the level of noise in the number of ribozymes or parasites in the RNA droplet experiment \cite{Matsumura2016}. 
We find from this table that the ribozyme size was $L=362$, 
and that the experiment should be in the replication-limited regime because the diffusion time scale
should be approximately over $2 \cdot 10^4$ times smaller than replication times of the order of $10$s.
The compositional variation due to noise in growth should be maximal when we start with one ribozyme and one parasite of equal length, and with $\alpha = \gamma$, which on average gives $\bar{x}=1/2$. Even then, the variation in final composition $\bar{x}$ is very small, since: $\sigma_{\bar{x}}\approx 0.02$. In such a case, our deterministic approach used in \cite{Blokhuis2018} is applicable.

\subsection{Phase diagram in the presence of weak noise}
\label{subsec: wn}

The growth equations given by Eqs. \eqref{mdet} and \eqref{ydet} are deterministic in nature, which means that a given initial condition $(n,m)$ yields a unique final composition $\bar{x}(n,m)$. 
In contrast to that in a stochastic approach, a given $n$ and $m$ lead to many different trajectories, which means 
that $\bar{x}(n,m)$ is a random variable with a probability distribution $p(\bar{x}(n,m))$. 
Consequently, the ribozyme fraction after one round is
\be
x'=\frac{\sum\limits_{n,m} \int^1_0 d \bar{x}  \bar{x}(n,m) p(\bar{x}(n,m))  f(\bar{x}) P_{\lambda}(n,x,m)}{\sum\limits_{n,m} \int^1_0 d \bar{x} p(\bar{x}(n,m))  f(\bar{x}) P_{\lambda}(n,x,m)}. \label{xprime2}
\ee
This expression is computationally demanding to evaluate for $\lambda \gg 1$, but it can be simplified significantly 
in the weak noise limit. 

\begin{figure}
\centering
\includegraphics[scale=0.40]{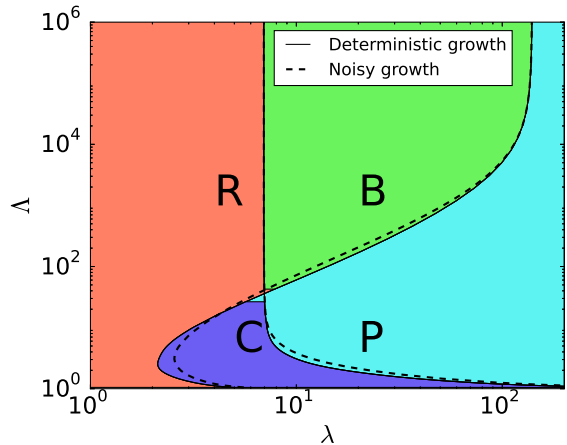} 
\caption{Phase diagram for ribozyme-parasite scenario in presence of noise given by Eq. \eqref{sigXX}, for $L_{\alpha}=L_{\gamma}=3$.}
\label{fig: phdiagN}
\end{figure}

In order to construct a phase diagram in this limit, we simplify Eq. \eqref{xprime2}, by considering $p(\bar{x}(n,m))\approx \mathcal{N}(\bar{x},\sigma_{\bar{x}})$, where $\mathcal{N}$ denotes a normal distribution with mean $\bar{x}$ and standard deviation defined by Eq. \eqref{sigXX}. 
From Eq. \eqref{sigXX} we expect the effect of noise to be largest when $\lambda, L$ and $\Lambda$ are close to 1 (if $\Lambda \gg 1, \bar{x} \rightarrow 0$).
In Fig. \ref{fig: phdiagN}, the original phase diagram from Ref. \cite{Blokhuis2018} is shown together with the modified phase boundaries (dotted lines) due to the presence of Gaussian noise using Eq. \eqref{xprime2} for the case that $L_{\alpha}=L_{\gamma}=3$. 

Given that the amplitude of this type of noise should rapidly diminish for larger $L$, and that $L \sim O(100)$ in the experiment, we expect our ribozyme-parasite scenario to be well-described by a deterministic dynamics.
We also see that the noise stabilizes the pure ribozyme phase (R) with respect to the coexistence phase (C) because in the presence of noise, the R region has grown at the expense of the C region. Similarly, the noise stabilizes the coexistence region (C) against the parasite region (P). 

\section{Conclusion}
\label{sec: conclusion}

In this paper, we have introduced two important extensions of our framework for transient compartmentalization
 \cite{Blokhuis2018}, by including the effect of noise in growth and deterministic mutations. These extensions
 have general implications for scenarios of the Origins of Life. In particular, we have shown that transient 
compartmentalization has a relaxed error threshold and is robust to noise. Although the former is also 
accomplished by the stochastic corrector model \cite{Szathmary1987}, transient compartmentalization is
 much more primitive: it needs neither cell division nor deterministic growth. 

In the presence of mutations, we have found that the phase diagram of this system only contains the parasite and the coexistence phases. The case where ribozymes grow faster than the parasites can be analyzed in terms of a modified error threshold, which interestingly now depends on the dynamics of compartmentalization and selection. The asymptotes of these error thresholds are derived analytically for arbitrary selection functions and allow to easily apply the theory to study various general classes of survival scenarios. 

To analyze the role of noise in growth, we have introduced a simple model for replication. The behavior of the model relies on the number of rate-limiting steps, which are assumed to be independent. For a single rate-limiting step, the model describes simple autocatalysis or rate-limiting binding to an enzyme. A polymer of length $L$ can have $L$ rate-limiting steps due to sequential monomer incorporation. When growing two simple autocatalysts competing for resources, there exists a fundamental trade-off: the closer the growth rates, the larger the compositional variability. Similar growth rates are important for coexistence, but compositional variability can be strongly detrimental to compartment survival. In the presence of selection, most compartments will acquire non-sustainable compositions. Selection mechanisms with cell division cannot afford such giant fluctuations, because at least half of the daughter cells need to survive to prevent extinction.  

Template replication can remove this problem. If it is rate-limiting, growth becomes increasingly deterministic with polymer size. As such, it will increase compartment survival. This may provide an evolutionary pressure pushing simple autocatalysts towards template replicators. Such a pressure would also enhance evolvability by making more powerful selection mechanism accessible to these replicators.

For the RNA droplet experiment \cite{Matsumura2016}, growth should be in the replication-limited regime, which we have quantified using tools borrowed from the theory of branching processes. In the weak noise limit, we constructed a modified phase diagram for our original ribozyme-parasite model. Transient compartments can use noise in growth to stabilize coexistence with a parasite.

Of course, the two effects that we have studied here separately, namely mutations and noise in growth, could be present simultaneously. We cannot also exclude that a more detailed modeling of the molecular 
replication or a different form of compartmentalization dynamics could lead to 
features not captured by the present treatment. Nevertheless, we think that the present 
framework represents a basis on which further studies could be built.
In particular, our results and their future extensions may have bearing on developments of important experimental techniques such as digital quantitative PCR \cite{Hindson2011} and Directed Evolution \cite{A.Drame-Maigne2018}. 

These findings invite us to consider the role of group selection during prebiotic evolution from a new perspective. 
A first question is whether transient compartmentalization could have allowed major transitions \cite{Smith1995} at the molecular level, 
such as the emergence of chromosomes \cite{Smith1993}, to occur before the advent of cell division. 
Second, we have shown that template polymerization, compared to autocatalysis of small molecules, enhances synchronization 
between compartmentalized replicators. 
This could have acted as a driving force for the transition from simple autocatalysts to more elaborate polymeric replicators. 
Third, the framework should be extended to integrate more elaborate growth dynamics. A recent extension\cite{Laurent2019} considers replicators which are themselves polymerases while parasites need the polymerase for their own growth. One of the main results of that study is that thanks to such interactions, there is no need of an explicit selection to preserve the replicases in a coexistence region by transient compartmentalization.
Finally, it remains to be elucidated how cell division and cell lineages could have emerged from transient compartments.

\section*{Acknowledgements}
A.B. was supported by the Agence Nationale de la Recherche 
(ANR-10-IDEX-0001-02, IRIS OCAV).
L.P. acknowledges support from a chair of the Labex CelTisPhysBio (ANR-10-LBX-0038).
We would like to thank Y. Rondelez for many important and insightful suggestions. We acknowledge stimulating discussions with B. Houchmandzadeh and within Institut de Convergences Qlife: 17-CONV-0005 Q-LIFE. We thank the referees for their helpful suggestions.

\appendix

\section{Population-level noise generated by a single individual in the initial condition}
\label{sec: A}

\noindent Let us consider an age-dependent renewal process, in which the probability density of branching at age $t$ is given by $f(t)$, and upon branching, the probability of having $k$ offspring is given by $\phi_{k}$ (assumed to be age-independent for simplicity). We would like to evaluate the behavior of the number $N(t)$ of individuals at time $t$. Let us define the function $h(s)$ by
\begin{equation}
h(s)=\sum_{k=0}^{\infty}\phi_{k}s^{k}.
\end{equation}
We also define the generating function for the process $N(t)$ by
\begin{equation}
G(s,t)=\sum_{k=0}^{\infty}p_{k}(t)s^{k},
\end{equation}
where $p_{k}(t)$ is the probability that $N(t)=k$. We assume that $p_{k}(0)=\delta_{k1}=p_{k}^{0}$, 
i.e., that we start from a single object. We can then evaluate $p_{k}(t)$ by adding the probability 
that no branching has occurred between 0 and~$t$, which is given by $1-\int_{0}^{t}\D t\;f(t)=1-F(t)$, 
with the effect of the first branching at time $u$, such that $0<u<t$. We obtain
\begin{equation}
p_{k}(t)=p_{k}^{0}\left(1-F(t)\right)+\sum_{l}\phi_{l}\int_{0}^{t}\D u\;f(u)\,\sum_
{\{n_{i}\}}\delta_{\sum_{i=1}^{\ell}n_{i},k}\prod_{i=1}^{\ell}p_{n_{i}}(t-u).
\end{equation}
Multiplying by $s^{k}$ and summing, we obtain
\begin{equation}
G(s,t)=s\left(1-F(t)\right)+\int_{0}^{t}\D u\;f(u)\,h\left(G(s,t-u)\right).
\end{equation}
Taking the derivative with respect to $s$ at $s=1$, we obtain the following equation for the average 
$\mu(t)=\sum_{k}k\,p_{k}(t)$:
\begin{equation}
\mu(t)=1-F(t)+m\int_{0}^{t}\D u\;\mu(t-u)\,f(u),
\end{equation}
where $m=h'(1)=\sum_{k}k\,\phi_{k}$ is the average number of daughters upon branching.

To solve this equation in the limit $t\to\infty$, let us multiply both sides by $\E^{-\alpha t}$ and take the limit. Since $\lim_{t\to\infty}F(t)=1$, we obtain
\begin{equation}
\begin{split}
\mu^{*}&=\lim_{t\to\infty}\mu(t)\E^{-\alpha t}=\lim_{t\to\infty}\int_{0}^{\infty}\D u\;m\mu(t-u)\,\E^{-\alpha (t-u)}\,\E^{-\alpha u}\,f(u)\\
&= \mu^{*}\,m\int_{0}^{\infty}\D u\;\E^{-\alpha u}f(u).
\end{split}
\end{equation}
This equation allows for a solution different from 0 and $\infty$ if $\alpha$ is chosen to satisfy
\begin{equation}
m\int_{0}^{\infty}\D u\;\E^{-\alpha u}f(u)=1.
\end{equation}
Then, making use of a result by Smith~\cite{Smith1953}, we obtain
\begin{equation}
\label{mu*}
\frac{1}{\mu^{*}}=\frac{\alpha m^{2}}{m-1}\int_{0}^{\infty}\D u\;u\,\E^{-\alpha u}f(u).
\end{equation}
As a consequence, we have
\begin{equation}
\mu(t)\approx \mu^{*}\,\E^{\alpha t}.
\end{equation}

In the case we are considering we have
\begin{equation}
f_{L}(t)=\frac{1}{\Gamma(L)}\,\kappa^{L}t^{L-1}\E^{-\kappa t},
\end{equation}
and $m=2$, which yields
\begin{equation}
\label{alpha}
\alpha=\kappa (2^{1/L}-1)\approx \frac{\kappa \ln 2}{L},
\end{equation}
giving, as long as $L\gg 1$,
\begin{equation}
\mu(t)\approx \frac{2^{\kappa t/L}}{2\ln 2}.
\end{equation}

We can use this framework to also evaluate higher moments of the population size, and
from that obtain the coefficient of variation of the population size which 
characterizes the amplitude of the noise.
Let us denote the second derivative of the generating function 
with respect to $s$ by $\zeta$
\be
\zeta(t)= \left. \frac{d^2 G(s,t)}{d^2 s}\right|_{s=1}=\sum^{\infty}_{k=1} (k (k-1)) p_k(t). 
\ee
At large times, $\zeta(t) \approx \zeta^* e^{2 \alpha t}$.
The variance of the population size $\sigma^2$ follows from the standard relation: 
\be
\sigma^2=\zeta + \mu - \mu^2 \simeq \zeta - \mu^2. \label{zettosig}
\ee
For the specific case we are considering, we find 
\be
\zeta^*= \frac{2 {\mu^*}^2}{(2^{\frac{L+1}{L}}-1)^L-2}.
\ee
After extracting the leading contribution in the large $L$ limit, we find:
\be
\frac{\sigma}{\mu} \approx \frac{\sqrt{2} \ln(2) }{\sqrt{L}}, \label{sqL1}
\ee
which is numerically close to $1/\sqrt{L}$ since $\sqrt{2} \ln 2 = 0.980.. \approx 1$.

\section{Population-level noise generated from $n$ individuals in the initial condition}
\label{sec: B}

If we start from $n$ individuals rather than just one, we can write the probability to have $k$ individuals at time $t$, $p^{(n)}_k(t)$, in terms of the subpopulations 
generated by $n$ single individuals, 
\begin{equation}
\begin{split}
p^{(n)}_{k}(t)&=\sum_{\{m_{1},\ldots,m_{n}\}}\delta_{\sum_{j}m_{j},k} \prod_{j=1}^{n} p^{(1)}_{m_{j}}(t). 
\end{split}
\end{equation}
Here, $p_k^{(1)}(t)$ denotes the probability of having a population size of $k$ at time $t$, 
starting from one individual, which was considered in \ref{sec: A}.
Note that we have added an addition superscript $(1)$ to the notation used in 
 \ref{sec: A} to emphasize the initial condition.
From this equation, the new generating function follows :
\begin{equation}
\begin{split}
G^{(n)}(s,t)=[G^{(1)}(s,t)]^n.
\end{split}
\end{equation}
From this equation, we obtain the average,
\begin{equation}
\mu^{(n)}(t)=n \mu^{(1)}(t) \label{mun}, 
\end{equation}
which expresses the average with $n$ initial strands in terms of the average with one initial strand. For the second moment, we obtain
\bea
\zeta^{(n)}=n(n-1) [\mu^{(1)}]^2 + n \zeta^{(1)},
\eea
We can then extract $\sigma^{(n)}$ by using Eq. \eqref{zettosig}, which yields
\bea
[\sigma^{(n)}]^2 &=& n (n-1) [\mu^{(1)}]^2 + n \mu^{(1)} - n^2 [\mu^{(1)}]^2 + n \zeta^{(1)} \\
&=& n \zeta^{(1)} + n \mu^{(1)} (1-\mu^{(1)}). \nonumber
\eea
Together with Eq. \eqref{mun}, this leads to
\be
\frac{\sigma^{(n)}}{\mu^{(n)}} \simeq \frac{\sqrt{\zeta^{(1)} - [\mu^{(1)}]^2}}{\sqrt{n}\mu^{(1)}}=
 \frac{\sigma^{(1)}}{\sqrt{n} \mu^{(1)}}
\ee
which is the coefficient of variation found previously for a single individual in the initial condition, divided by $\sqrt{n}$
as expected for the growth from independent individuals.
This confirms the scaling found in Eq. \eqref{sqL_fin}.
 
\bibliography{JTBbib.bib}

\end{document}